\documentstyle[twocolumn,prl,aps,epsfig]{revtex}
\begin{document} 
\draft 
\title{Helium Clustering in Neutron-Rich Be Isotopes}
\author{N.I.~Ashwood$^1$, M.~Freer$^1$, S.~Ahmed$^1$, J.C.~Ang\'{e}lique$^2$, V.~Bouchat$^3$, W.N.~Catford$^{4,2}$,
N.M.~Clarke$^1$, N.~Curtis$^1$, O.~Dorvaux$^5$, B.R.~Fulton$^6$, F.~Hanappe$^3$, Y.~Kerckx$^3$,
M.~Labiche$^7$, J.L.~Lecovey$^2$$^{\dagger}$, R.C~Lemmon$^8$, F.M.~Marqu\'{e}s$^2$, T.~Materna$^3$,
A.~Ninane$^9$, G.~Normand$^2$, N.A.~Orr$^2$, S.~Pain$^4$, N.~Soi\'{c}$^1$$^{\ddagger}$, L.~Stuttg\'{e}$^5$,
C.~Timis$^2$$^*$, A.~Unsharova$^{10}$, J.S.~Winfield$^{11}$ and V.A.~Ziman$^1$}

\address{
$^1$School of Physics and Astronomy, University of Birmingham, Edgbaston, Birmingham, B15 2TT, United Kingdom.\\
$^2$Laboratoire de Physique Corpusculaire, ISMRA and Universit\'{e} de Caen, IN2P3-CNRS, 14050 Caen Cedex, France.\\
$^3$Universit\'{e} Libre de Bruxelles, CP 226, B-1050 Bruxelles, Belgium.\\
$^4$School of Electronics and Physical Sciences, University of Surrey, Guildford, Surrey, GU2 7XH, United Kingdom.\\
$^5$Institut de Recherches Subatomique, IN2P3-CNRS/Universit\'{e} Louis Pasteur, BP 28, 67037 Strasbourg Cedex, France.\\
$^6$Department of Physics, University of York, Heslington, York, YO10 5DD, United Kingdom.\\
$^7$Department of Electronic Engineering and Physics, University of Paisley, Paisley, PA1 2BE, United Kingdom.\\
$^8$CLRC Daresbury Laboratory, Daresbury, Warrington, Cheshire, WA4 4AD, United Kingdom.\\
$^9$Institut de Physique, Universit\'{e} Catholique de Louvain, Louvain-la-Neuve, Belgium.\\
$^{10}$Joint Institute for Nuclear Research, 141980, Dubna, Moscow Region, Russia.\\
$^{11}$ Instituto Nazionale di Fisica Nucleare, Laboratori Nazionali del Sud, 44 - 95123 Catania, Italy.\\}

\date{Received \today}
\maketitle
 
\begin{abstract} 
Measurements of the helium-cluster breakup and neutron removal 
cross sections for neutron-rich Be isotopes $^{10-12,14}$Be are 
presented. These have been studied in the 30 to 42 MeV/u energy 
range where reaction measurements are proposed to be sensitive to 
the cluster content of the ground-state wave-function. These 
measurements provide a comprehensive survey of the decay processes 
of the Be isotopes by which the valence neutrons are removed 
revealing the underlying $\alpha$-$\alpha$ core-cluster structure. 
The measurements indicate that clustering in the Be isotopes 
remains important up to the drip-line nucleus $^{14}$Be and that 
the dominant helium-cluster structure in the neutron-rich Be 
isotopes corresponds to $\alpha$-$Xn$-$\alpha$. 
\end{abstract} 
 

The structure of nuclear matter at the limits of stability has
always been a source of fascination. These limits are bounded by
the extremes of excitation energy or spin, on one hand, and by the
extremes of either total number of nucleons or ratios of protons
to neutrons (isospin) on the other. Most notably, radioactive beam
facilities provide access to the limit of neutron excess. Here,
structural properties are known to differ from normal nuclear
matter, as they are strongly influenced by the weak binding of the
valence neutrons. The classic example is the halo, where weakly
bound neutrons exist as a diffuse cloud around a compact core
\cite{rev4}. In this case there is a decoupling of the valence
neutrons from a core, or cluster, of normal nuclear matter.

Recently, it has been speculated that such clustering may be a
general feature of drip-line nuclei caused by the tendency of the
nucleus to distribute the neutron excess amoung the core nucleons
\cite{Hor00}. In order to maximize the overlap of the excess
neutrons with protons, the core may either deform
 enhancing the surface area, or optimally, form clusters. Thus, the
ground states of neutron drip-line nuclei may resemble clusters
embedded in a cloud of weakly bound neutrons.

Certain nuclei already possess a predisposition towards
clustering. For example, $^8$Be clearly demonstrates a strong
$\alpha$-$\alpha$ cluster character \cite{Buc77,Hiu72}. Recent
experimental and theoretical developments indicate that this
underlying cluster structure has a strong influence on the
neutron-rich Be isotopes. In particular, the two-centred nature of
the potential in which the valence neutrons reside may produce
molecular type characteristics in these nuclei \cite{von97,Ita00}.
Calculations using the antisymmeterized molecular dynamics (AMD)
framework suggest such features can occur in the Be isotopes
\cite{Kan99,Kan95}. The same model predicts that the normally
compact boron isotopes become highly clustered at the drip-line
\cite{Kan95a}, in line with the ideas concerning the optimization
of the proton valence-neutron overlap \cite{Hor00}.

Direct access to such structural properties of the ground states
of very neutron-rich nuclei is extremely challenging due to the
fact that the $^x$He cluster breakup thresholds increase
significantly towards the drip-line. However, the AMD model has
been used to explore how the fragmentation of beams of boron
isotopes may be used to extract a signature for the clustering in
the ground state \cite{Tak01}. This study revealed that such
clustering may be probed via fragmentation if energies in the
region of 30 MeV/nucleon were used. However, studies of the most
neutron-rich boron isotope ($^{19}$B), as suggested in
\cite{Tak01}, is beyond the capability of present day experimental
facilities. It is, however, possible to study the beryllium nuclei
all the way from stability to the drip-lines. Here we present
comprehensive measurements of $^x$He-cluster, which were explored
theoretically for the boron isotopes, and neutron breakup cross
sections for the beryllium isotopes A=10, 11, 12 and 14. These
data should provide a significant test of our understanding of the
structure of nuclei at the neutron drip-line.

The measurements were performed at the GANIL accelerator facility
with the production of $^{10,11,12,14}$Be secondary beams via the
fragmentation of $^{13}$C and $^{18}$O primary beams. The reaction
products were mass, charge and momentum analyzed using the LISE3
spectrometer producing beams of purity $\ge$95$\%$ for the lighter
isotopes and $\simeq$20$\%$ for the heaviest and count rates of
the order of 10$^4$ pps for masses 10, 11 and 12 (limited by the
count rate capacity of the detection system) and $\sim$50 pps for
$^{14}$Be. Identification of the beam of the beam particle was
achieved using time-of-flight through the LISE separator as
measured by a PPAC at the entrance to the reaction chamber. The
energies of the 4 beams were 30.9, 41.7, 41.8 and 34.4 MeV/u in
order of increasing mass.

The beam was tracked onto 20 to 275 mg/cm$^{2}$ carbon targets
using two drift chambers, providing a measurement of the incident
position and angle of the projectile. The beam and reaction
products then entered a zero-degree telescope formed from two 500
$\mu$m thick, 16-strip position-sensitive silicon detectors placed
15 cm from the target. These two detectors were arranged with
orthogonal strips, providing a measurement of the incident ions to
$\le$1 mm in both the $x$ and $y$ directions (the $z$-coordinate
being the beam direction). Behind the strip detectors was a close
packed array of 16 1.5 cm thick, 2.5$\times$2.5 cm$^2$, CsI
scintillators. These were placed at 30 cm from the target so as to
cover the same solid angle as the strip detectors as measured from
the target. Calibration of the energy response of the silicon and
CsI detectors was achieved using $\alpha$-sources and a mixed beam
of light-ions of known energy. Neutrons produced in reactions of
the beam were detected using an array of $\sim$100 neutron
detectors (D\'{e}MoN) arranged in a similar manner to that shown
in \cite{Mar02}. These detectors gave a single-neutron detection
efficiency of $\sim$15 $\%$ (as in \cite{Lab01}). Monte Carlo
simulations of the response of the charged particle detection
system indicates that the efficiency for the detection of the
breakup of the Be isotopes into two $^x$He nuclei was 40 to 50
$\%$ and almost independent of excitation energy of the decaying
system. The measured relative energy of the decay products
($E_{rel}=\mu v_{rel}^2$/2) thus allowed the excitation energy to
be deduced ($E_x=E_{rel}+E_{thresh}$).

A knowledge of the breakup yields, the target thickness, detection
efficiency and the integrated beam exposure allowed the cross
sections for the various breakup processes to be calculated. In
addition, as noted above, it was possible to deduce the excitation
energy of the decaying system. As an example, Fig. 1 shows the
$^4$He+$^4$He invariant mass spectrum ($E_x$) for the $^{12,14}$Be
beams. It is clear that a large fraction of the events involving
the dissociation of $^{14}$Be ($^{12}$Be) proceed via 6 (4)
neutron emission to form $^8$Be, i.e. the decay process does not
always follow from the decay into neutron-rich helium isotopes
followed by neutron emission, i.e. unbound excited states of
$^x$He. The evidence for this is the observation of peaks in the
$^8$Be excitation energy spectrum at energies which correspond to
the decay of the ground-state (92 keV) and the first excited state
(3.04 MeV, 2$^+$). There is one further feature in this spectrum,
at $\sim$600 keV, which does not correspond to a resonance in
$^8$Be but is produced by the decay of the 2.43 MeV (5/2$^-$)
state in $^9$Be to the tail of the $2^+$ state in $^8$Be. The
presence of this feature suggests a fraction of the yield proceeds
via sequential 6n emission to $^8$Be. Similar features are also
present in the $^4$He+$^4$He decay spectra measured with the other
Be projectiles.

The mass identification provided by the zero degree telescope
allows the resolution of all the He isotopes and thus permits all
stages of the decay process to be reconstructed. For example, in
the case of $^{14}$Be it is possible to reconstruct the decay into
$^6$He+$^8$He, or following the emission of neutrons from the
projectile, the decay of $^{12}$Be$^*$ into $^6$He+$^6$He (or
$^4$He+$^8$He), or the decay of $^{10}$Be$^*$ to $^6$He+$^4$He, or
finally, as discussed above, $^8$Be to $^4$He+$^4$He. The cross
sections for these processes are plotted in Fig. 2. These cross
sections are also presented together with the $^x$Be+1n (i.e. core
fragment plus one neutron) cross sections in Table 1. The neutron
cross sections have been calculated by selecting beam velocity
neutrons ($E_n>$0.37E$_{beam}$/A). In the case of $^{11,12}$Be
these were calculated by integrating the neutron angular
distributions with a Lorentzian single neutron line-shape, whilst
subtracting the background from reactions within the CsI component
of the telescopes, as deduced from measurements with no target
\cite{Lab01}. In the case of $^{10,14}$Be data no background
measurements were performed and thus the cross sections provides
upper limits only. For $^{14}$Be, measurements of these cross
sections, including the background subtraction, have already been
made by Labiche {\em et al.}\cite{Lab99,Lab01} at 35 MeV/u, and
these are also shown in Table 1. Given an analysis of the no
target yield from the $^{12}$Be data ($\sim$20-30$\%$ of events in
2n removal channel are from the target) the cross sections are
consistent. Comparison of the cross sections with earlier studies
show reasonable agreement. The 0.18$\pm$0.11 b $^{11}$Be 1n
removal cross section compares with that extracted by Anne {\em et
al.} \cite{Ann94} of 0.12$\pm$02 b, and the 0.11$\pm$0.04 b
$^{12}$Be 1n removal is in close agreement with that measured by
Navin {\em et al.} \cite{Nav00}, 0.045$\pm$0.05 b, albeit at 78
MeV/u. Similar agreement is found with the helium breakup
measurements reported in \cite{Fre01}.

The measurement of the multi-neutron removal channels via the
detection of the core plus one neutron presents the advantage that the
detection efficiency does not become prohibitively small. Naively, the
true cross section for the $A-Xn$ channel should be reduced by a
factor $X$. However, the neutron angular distributions vary from decay
step to decay step and thus the efficiency is not constant, and,
moreover, the multiplicity of the emitted neutrons does not
necessarily equal the number of missing neutrons, with the possibility
of projectile neutrons interacting strongly with the target
\cite{Lab01}. Indeed, the measured neutron multiplicities for
$^{14}$Be$\rightarrow^{12}$Be and $^{14}$Be$\rightarrow^{10}$Be are
found to be 1.63$\pm$0.26 and 2.9$\pm$0.8 \cite{Lab99,Lab01}. Although
the error bars are large these multiplicities indicate that there is a
tendency for there to be slightly less neutrons in the final state
than anticipated.  Nevertheless, for the sake of the comparison with
the measured helium breakup cross sections the $^{A-X}$Be+1n cross
sections are plotted in Fig. 2 divided by $X$, thus indicating the
strength with which the bound states of the $A-Xn$ nucleus are
populated.

It is clear that neutron removal dominates even in the case of the
nucleus $^{10}$Be in which the neutron decay threshold (6.8 MeV)
is almost commensurate with that for $\alpha$-decay (7.4 MeV). In
all cases the breakup of the projectile into helium fragments is
several orders of magnitude smaller than the neutron removal cross
sections. The helium breakup cross sections steadily increase as
the neutron removal process feeds states above the cluster decay
thresholds. At the point at which the systems arrive at $^8$Be the
cluster breakup and neutron removal processes coincide. For each
step the ratio of yield in the $A-Xn$ and helium breakup gives an
idea of the relative excitation probabilities to states above the
neutron and cluster decay thresholds, and subsequent stages reveal
the probability that the daughter $A-Xn$ system is formed in
either bound states or states unbound to either neutron of helium
decay. These decay systematics should allow detailed modeling of
the decay process via which these nuclei lose all of their
neutrons leaving the valence $\alpha$-$\alpha$ cluster core.

Fig. 3 shows the mass dependence of the cluster decay cross
sections. It is not possible to plot the data for $^{11}$Be in
this instance as the main decay channels $^{5}$He+$^{6}$He
$^{4}$He+$^{7}$He lead to a $^{4}$He+n+$^{6}$He final state which
cannot be unambiguously distinguished from neutron removal
followed by $\alpha$-decay of $^{10}$Be. Figs 3a and 3b show the
cross sections for the helium cluster breakup of the projectile
(first-chance breakup), i.e. $^{14}$Be$\rightarrow^{6}$He+$^8$He,
$^{12}$Be$\rightarrow^{6}$He+$^6$He,
$^{10}$Be$\rightarrow^{6}$He+$^4$He, and the total He+He breakup
cross section respectively. The first chance breakup cross section
is essentially mass independent. These systematics indicate that
the di-cluster content of the nuclei remains relatively constant
from $^{10}$Be through to $^{14}$Be. However, the total breakup
cross section for A=10, 11, 12 and 14 shows a strong minimum at
A=11 increasing to maximum values for $^{10}$Be and $^{14}$Be. The
ratio of first chance to total cluster breakup thus reveals a
maximum at A=12.

Early calculations of the degree of clustering in the beryllium
isotopes using the AMD framework indicated a decrease in
clusterization from $^8$Be to $^{12}$Be with a small increase for
$^{14}$Be \cite{Kan95}, with something like a factor of 2 change
predicted in the clusterization between the ground states of
$^{10}$Be and $^{12}$Be. However, most recent studies using this
framework now indicate low-lying states posses strong cluster
symmetries \cite{Kan03}. If, as suggested by the AMD calculations
for the reactions with the boron isotopes, the breakup process is
strongly influenced by the overlap of the cluster content of the
ground state and that of excited states above the decay threshold,
then the present measurements indicate that there is no change in
the degree of clustering in the mass range 10 to 14.

By far the most dominant He-cluster breakup channel is that
corresponding to the removal of all of the valence neutrons
leaving the $\alpha$-$\alpha$ core (Fig 2). Table 2 shows the
energy thresholds for $^8$Be+$Xn$ and first-chance He-cluster
breakup. Remarkably, even though the thresholds for removing all
valence neutrons are higher in energy, it is this channel which
dominates, suggesting a strong influence of the
$\alpha$-$Xn$-$\alpha$ structure. Such a result cannot be
explained by decay phase-space alone or the fact that with high
numbers of valence neutrons there are many more partitions for the
system to decay into. For example, in the case of $^{14}$Be the
decay threshold for 6n emission lies 4.2 MeV higher than that for
cluster decay, and inelastic excitation probabilities would be
expected to decrease with excitation energy. The $^{14}$Be nucleus
must emit one of six possible neutrons to resonant states above
the $^{13}$Be 5n decay threshold (10.3 MeV), rather than to lower
energy states. In turn $^{13}$Be must decay to resonances above
the $^{12}$Be 4n decay threshold. Thus, although there is a factor
of 6 enhancement in the number of possible decay paths for the
neutron emission, there are 5 intermediate steps in which the
system may decay to states below the $\alpha$+$\alpha$+xn
threshold. Moreover, the phase space at each step is highly
constrained.

It is possible that the excitation mechanisms leading to the
first-chance He-cluster breakup and neutron removal are radically
different. Certainly, direct elastic breakup and absorptive
processes dominate in the first step over resonant breakup as they
have been used extensively to determine the ground state structure
of neutron-rich nuclei \cite{rev4}. It is thus probable that a
significant fraction of the $\alpha$+$\alpha$+6n yield for
$^{14}$Be is produced via such processes.

The measurements presented here provide evidence for the existence
of di-cluster structures in $^{10-12,14}$Be. Moreover, they appear
to demonstrate that these nuclei possess a strong structural
overlap with an $\alpha$-$Xn$-$\alpha$ configuration. The
comprehensive measurements of the neutron-removal and cluster
breakup for the first time provide experimental data whereby the
structure of the most neutron-rich Be isotopes can be modeled via
their reactions.

The authors are grateful to the technical and operations staff of
LPC and GANIL. This work was funded by the EPSRC (UK) and the
IN2P3-CNRS (France). Additional support was provided by the Human
Capital and Mobility Programme of the European Community (Contract
no. CHGE-CT94-0056).

\newpage
$.$
\newpage
\footnotesize
\begin{tabular}{|l||c|c|c|c||c|c|c|c|c|c|c|c|c||c|}\hline\hline
 & 1n & 2n & 3n & 4n  & $^6$He+$^8$He & $^6$He+$^6$He & $^8$He+$^4$He & $^6$He+$^4$He & \multicolumn{3}{c|}{$^4$He+$^4$He} & Total  \\
 &   & & & & &                &               &               &  $^8$Be$_{gs}$ & $^9$Be(5/2$^-$)  & $^8$Be(2$^+$) & He+He\\ \hline
$^{14}$Be  & & 2248(82) & 670(23) & 799(26)  & 1.17(0.20) &
1.33(0.22) & 2.15(0.30) & 10.04(0.65) &
   4.40(0.41) & 5.40(0.44) & 16.61(1.19) & 41.09(1.54) \\
$^{14}$Be$^\dagger$  & & 750(10) & & 420(10) &    &  & & & & & & \\
$^{12}$Be & 113(43) & 237(54) & 100(35)& & &  0.23(0.02) & 1.02(0.04) & 3.04(0.05) &
  2.89(0.10) & 4.25(0.10) & 16.42(0.37) & 27.86(0.40) \\
 $^{11}$Be & 175(110)   & 22(12) &  &  & & & & 1.17(0.07) &
   3.59(0.14) & 4.52(0.11) & 13.12(0.31) & 22.40(0.36) \\
 $^{10}$Be & 29(4)   & & & & & & &  1.26(0.06) &

   7.97(0.40) & 7.32(0.30) & 21.6(1.1) & 38.15(1.36) \\
\hline\hline
\end{tabular}
\vspace{0.5 cm}
\begin{tabular}{|c|cccccc|cc|}
\hline\hline & \multicolumn{6}{c|}{Neutron removal} &
\multicolumn{2}{c|}{He+He cluster}
\\& 1n & 2n & 3n & 4n & 5n & 6n & decay & \\ \hline
$^{14}$Be & 2.97 & 1.12 & 4.29 & 4.79
&11.60 & 13.27 & $^{6}$He+$^{8}$He & 9.09
\\
$^{12}$Be & 3.17 & 3.67 & 10.48 & 12.15 & &  & $^{4}$He+$^{8}$He & 8.95 \\
$^{11}$Be & 0.50 & 7.32 & 8.98 &  & &  & $^{5}$He+$^{6}$He & 8.81 \\ $^{10}$Be
& 6.81 & 8.48 &  &  & &  & $^{4}$He+$^{6}$He & 7.41 \\ \hline \hline
\end{tabular}
\normalsize
\vspace{0.5 cm}
{\bf Figure and Table Captions}\\
{\bf Table 1.} Cross sections measured for He-cluster breakup and
core plus one neutron in units of mb. Note in the case of
$^{10,14}$Be the neutron cross sections provide upper limits. The
row marked $^{14}$Be$^\dagger$ are the background subtracted cross sections from \cite{Lab99,Lab01}.\\
{\bf Table 2.} Energy thresholds for cluster and neutron decay in units of

MeV.\\
{\bf Figure 1.} The excitation energy spectrum corresponding to
$\alpha$+$\alpha$ coincidences from the $^{12}$Be beam. The peaks
correspond to the the decay of the ground-state (92 keV) and the
first excited state (3.04 MeV, 2$^+$). The peak at $\sim$600 keV
is produced by the decay of the 2.43 MeV (5/2$^-$) state in $^9$Be
to the tail of the $2^+$ state in $^8$Be. The inset is the
excitation energy spectrum corresponding to
$\alpha$+$\alpha$ coincidences from the $^{14}$Be beam.\\
{\bf Figure 2.} Normalized neutron-removal (open diamonds) and
cluster breakup (filled circles) cross sections for the
neutron-rich Be isotopes. For $^{14}$Be the filled diamonds
correspond to the measurements of \cite{Lab99,Lab01}. Note, the
points for the $^9$Be breakup (open circles) are lower limits
deduced from the 600 keV
peak in the $^8$Be spectrum in Figure 1. The vertical dotted lines indicate the mass
of the projectile.\\
{\bf Figure 3.} (a) The first-chance cluster breakup cross
sections, i.e. $^{14}$Be$\rightarrow^{6}$He+$^{8}$He,
$^{12}$Be$\rightarrow(^{6}$He+$^{6}$He and  $^{4}$He+$^{8}$He) and
$^{10}$Be$\rightarrow^{6}$He+$^{4}$He, (b) the total He+He breakup cross sections.\\

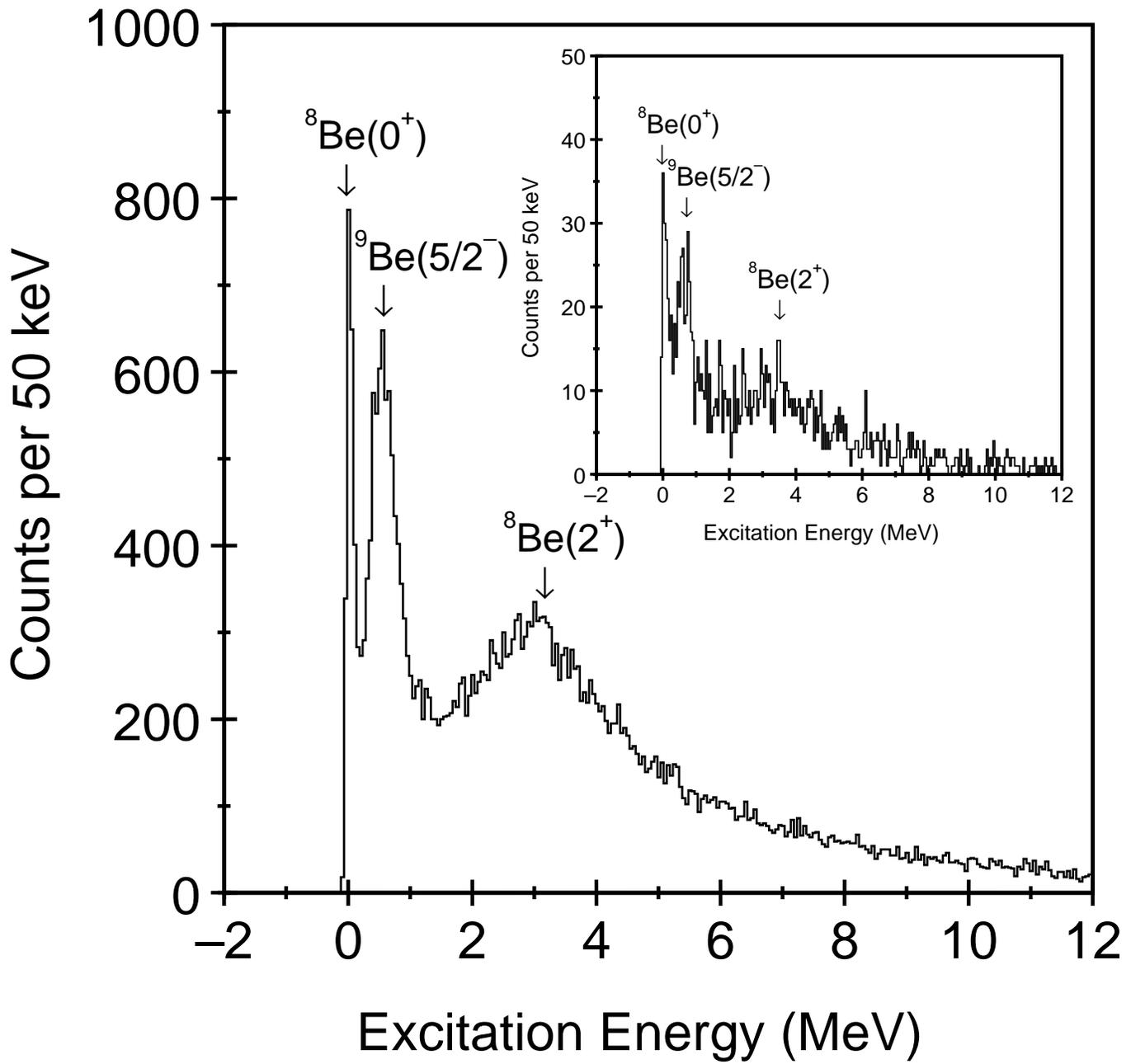
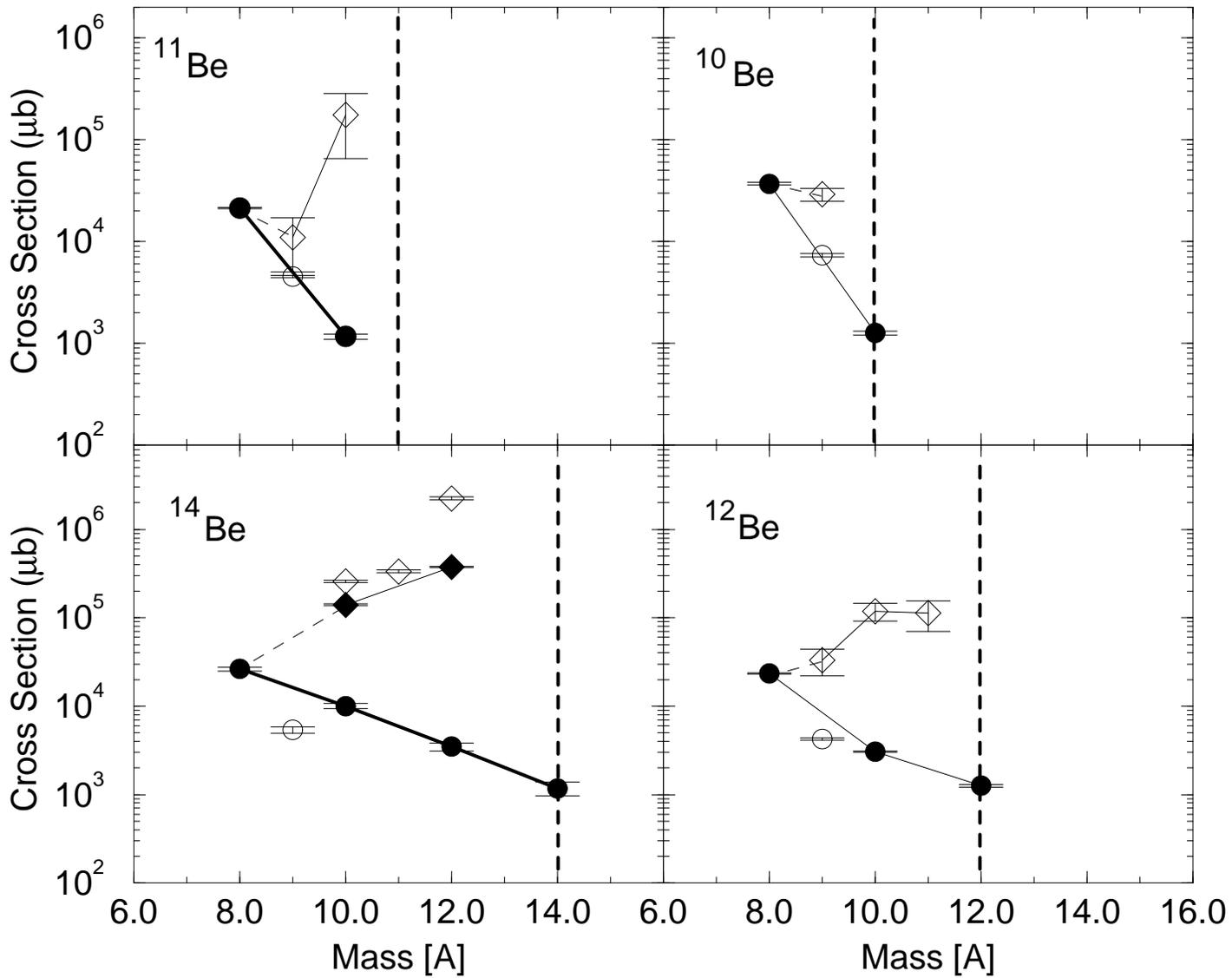
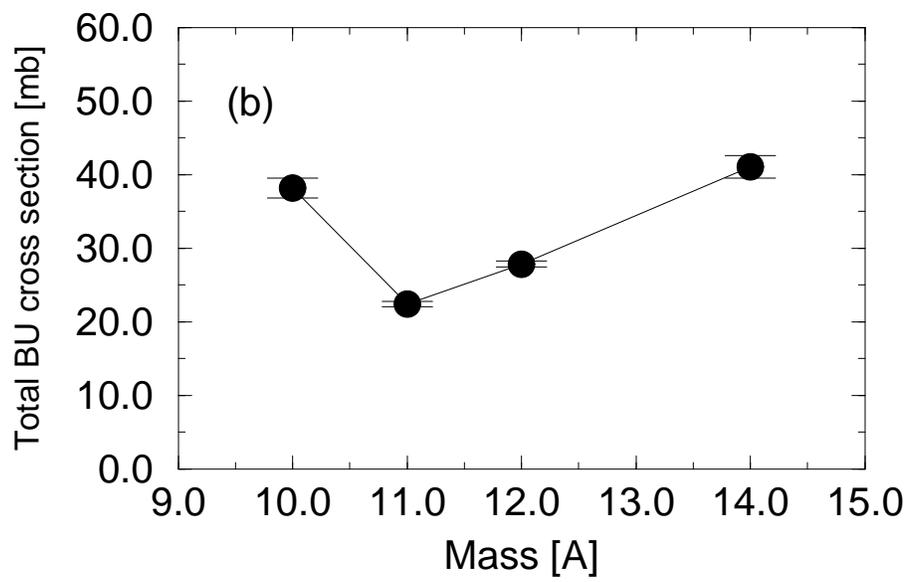
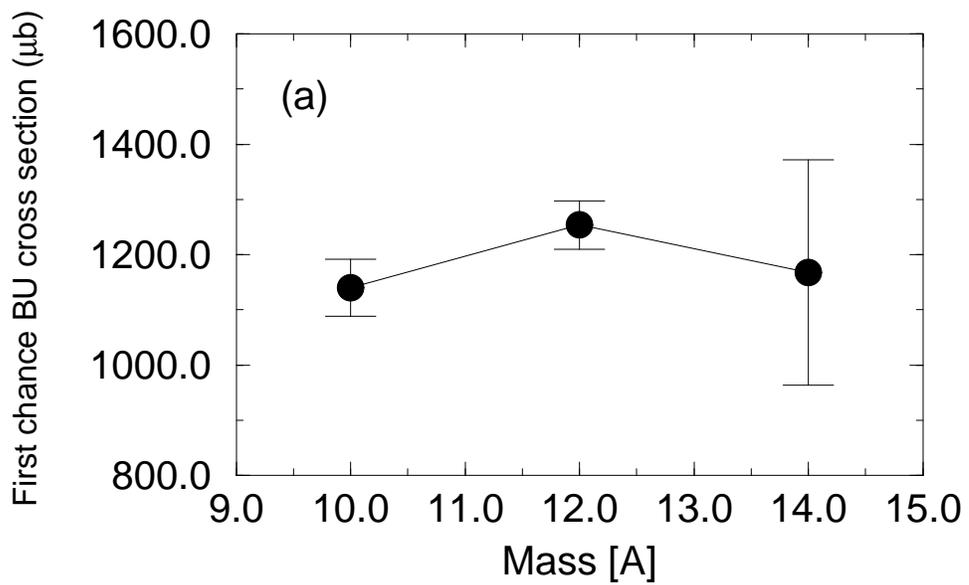
\end{document}